\documentclass[%
 apl,%
 amsmath,%
 amssymb,%
 amsfont,%
 preprint,%
]{revtex4-1}

\usepackage{graphicx}
\usepackage{bm}
\usepackage{color}

\usepackage{natbib}

\begin{document}

\title{ Strain effects on the thermal properties of ultra-scaled Si nanowires }%
\author{Abhijeet Paul}
\email{abhijeet.rama@gmail.com}
\author{Gerhard Klimeck}

\affiliation{School of Electrical and Computer Engineering, Network for Computational Nanotechnology, Purdue University,%
 West Lafayette, Indiana, USA, 47907.}

\pagenumbering{arabic}

\date{\today}

\begin{abstract}


The impact of uniaxial and hydrostatic stress on  the ballistic thermal conductance ($\kappa_{l}$)  and the specific heat ($C_{v}$) of [100] and [110] Si nanowires are explored using a Modified Valence Force Field phonon model. An anisotropic behavior of $\kappa_{l}$ and isotropic nature of $C_{v}$ under strain are predicted for the two wire orientations. Compressive (tensile) strain decreases (increases) $C_{v}$.
The $C_{v}$ trend with strain is controlled by the high energy phonon sub-bands. Dominant contribution of the low/mid (low/high) energy bands in [100] ([110]) wire and their variation under strain governs the behavior of $\kappa_{l}$.  


\end{abstract}


\maketitle 

The wide application of Silicon nanowires (SiNWs) in areas ranging from Complementary-Metal-Oxide-Semiconductor (CMOS) transistors \cite{Sinwfet,sinwfet2} to thermoelectric (TE) devices \cite{Hochbaum2008}, non-volatile memories \cite{sinw_nvm} and, solar cells \cite{sinw_solar_cells} has made SiNWs an extremely useful nano-electronics device concept. 
Device properties can be engineered \cite{sinw_electron_strain_1,sinw_electron_strain_2} through the introduction of built-in stress by process engineering and by external forces like wafer bending \cite{strain_size_sinw}. Tuning thermal properties using strain can be beneficial for cooling and increasing gain of lasers \cite{strain_size_sinw,strain_effect_1} or improving the efficiency of TE devices \cite{strain_effect_1}.

\textit{Prior Work:} Strain/stress effects on the thermal conductivity of doped bulk zinc-blende (ZB) semiconductors at low temperatures has been well studied \cite{phonon_uniax,Stress_phonon_GaAs}. Most of the bulk studies in ZB semiconductors found decrease in thermal conductivity with tensile strain which is qualitatively attributed to (i) a decrease in the phonon mean free path \cite{thermal_inc_quality}, (ii) an increase in the Debye temperature \cite{strain_size_sinw,thermal_inc_quality}, and  (iii) a change in the material stiffness \cite{shape_depend_theory}. The few experimental \cite{sige_strain_measure} and theoretical  \cite{strain_size_sinw,strain_effect_1,strain_effect_2,shape_depend_theory} efforts that have focused on the strain effects on the thermal properties in nanostructures, are either limited to single crystal orientation (mostly [100]) or diffusive phonon transport even in small nanostructures.  This work investigates the external strain effects on the thermal properties in ultra-scaled SiNWs in the coherent phonon transport regime where the wire cross-section sizes are comparable to the phonon wavelengths \cite{coherent_phonon} ($\lambda_{ph} = hV_{snd}/k_{B}T_{300K}$ $\approx 1nm$  at $V_{snd}$ = 6.5 km/sec for a 3nm X 3nm [100] SiNW \cite{vff_iwce_paper}, where $k_{B}$ and h are the Boltzmann constant and Planck's constant, respectively).

Free-standing SiNWs' phonon dispersion are studied using the modified valence force (MVFF) model \cite{VFF_mod_herman,vff_own_paper,vff_iwce_paper}. The MVFF model has been shown to correctly reproduce the strain effects like the Gruneisen parameters \cite{VFF_mod_herman,vff_own_paper} and third order elastic constants in bulk Si and Ge \cite{VFF_mod_herman}. The same set of Si MVFF parameters \cite{VFF_mod_herman,vff_own_paper} are used in this work to model the SiNW phonon dispersion under hydrostatic and uniaxial stress.

The ballistic thermal conductance ($\kappa_{l}$) across a semiconductor wire maintained under a small temperature gradient $\Delta T$, can be calculated from the calculated phonon dispersion, using the Landauer's formula \cite{Land} using Eq.(\ref{eq_kappa}) \cite{jauho_method,mingo_ph},
\begin{footnotesize}
\begin{equation}
\label{eq_kappa}
 	\kappa_{l}(\epsilon,T) = \frac{e^2}{\hbar} \sum_{i\in[1,N]} \kappa^{i}_{l}(\epsilon,T),\;\text{where}
\end{equation}

\begin{equation}
\label{eq_FE}
 	\kappa^{i}_{l}(\epsilon,T) = \int^{E_{i,max}}_{E_{i,min}} M(\epsilon,E_{i})\cdot E_{i} \cdot \frac{\partial}{\partial T}\Big[\frac{1}{(exp\big(\frac{E_{i}}{k_{B}T})-1\big)} \Big]dE_{i} 
\end{equation}
\end{footnotesize}

where N is the number of energy bins in the entire phonon dispersion energy range.The terms $\epsilon$, M(E), T, and  $e$ are the strain percentage, the number of modes at phonon energy E, the temperature, and the electronic charge, respectively. The term $\kappa^{i}_{l}(E,T)$ is the contribution to the total $\kappa_{l}$ of SiNW for $E_{i,min}\le$ E $<E_{i,max}$. This energy resolved representation in Eq. (\ref{eq_kappa}) allows for better understanding of the variation in $\kappa_{l}$ due to strain.

From the calculated phonon dispersion the constant volume specific heat ($C_{v}$) at a given T, can be calculated \cite{Fph_1} as,
\begin{footnotesize} 
\begin{eqnarray}
\label{cv_eqn}
	C_{v}(T) & = & (\frac{k_{B}}{m_{uc}})\cdot\sum_{n,q} \Big[ \frac{ (\frac{E_{n,q}}{k_{B}T})\cdot \exp(\frac{-E_{n,q}}{k_{B}T})}{[1-\exp(\frac{-E_{n,q}}{k_{B}T})]^{2}} \Big] \quad [J/kg.K],
\end{eqnarray}
\end{footnotesize} 
where $m_{uc}$ is the mass of the SiNW unitcell in kg. The quantity $E_{n,q}$ is the phonon eigen energy associated with the branch `n' and crystal momentum vector `q'. 

\begin{figure}[hbt!]
	\centering
		\includegraphics[width=3.2in,height=1.7in]{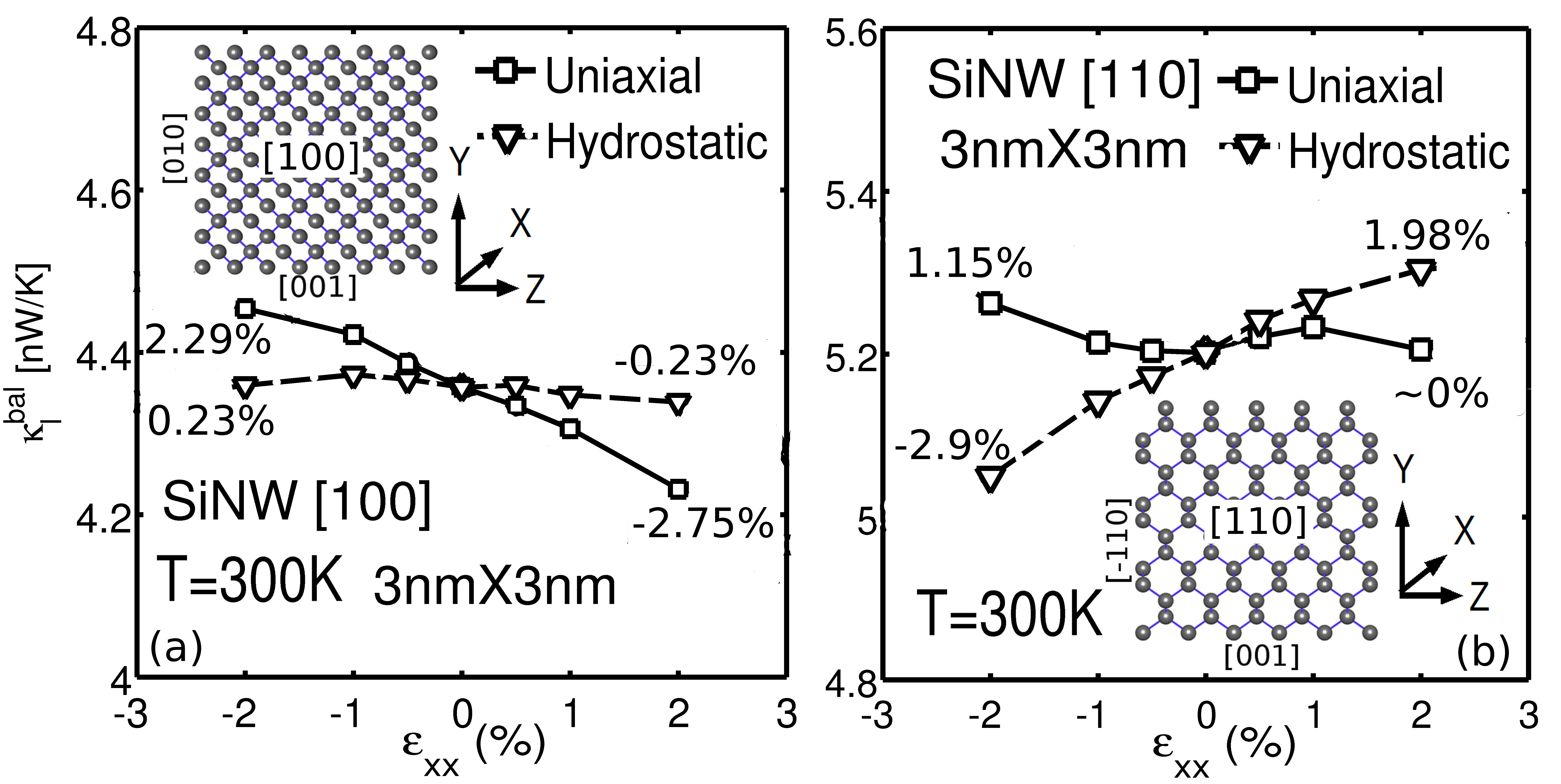}
	\caption{Variation in $\kappa_{l}$ with hydrostatic and uniaxial stress in 3nm $\times$ 3nm SiNW with (a) [100] and (b) [110] orientation. The uniaxial stress is applied along the wire axis. Insets also show the atomic structures of both the SiNWs with the coordinate axes.}
	\label{fig:sinw_thermal_cond}
\end{figure}

\textit{SiNW details:} Square NWs with width (W) and height (H) = 3nm with two channel orientations of $[$100$]$ and $[$110$]$ are studied (inset of Fig. \ref{fig:sinw_thermal_cond}a,b). Hydrostatic deformation (equal deformation along all the directions) and uniaxial deformation (along the wire axis) are applied to these wires varying up to $\pm$2\%. The outer surface atoms in these NWs are allowed to vibrate freely. Extremely small SiNWs may show significant surface and internal atomic reconstruction \cite{Amrit,strain_effect_1} or phase change \cite{strain_effect_1} under strain leading to larger changes in $C_{v}$ and $\kappa_{l}$, are outside the scope of the present study.   


\textit{Ballistic thermal conductance $\kappa_{l}$:} In all the SiNWs $\kappa_{l}$ is calculated using Eq.(\ref{eq_kappa}) at 300K. $\kappa_{l}$ increases (decreases) under uniaxial compression (tension) for both the wire orientations. Similar variations in $\kappa_{l}$ for [100] SiNWs are also observed by Li. et al\cite{strain_effect_1} using non-equilibrium molecular dynamics (NEMD) calculations. 
The [100] SiNW shows larger variation in $\kappa_{l}$ under tensile uniaxial deformation compared to the [110] SiNW (Fig. \ref{fig:sinw_thermal_cond}a,b). $\kappa_{l}$ shows a weak hydrostatic stress dependence in [100] SiNW in contrast to the [110] SiNW which shows a decrease (increase) of 2.9\% (1.98\%) in $\kappa_{l}$ under 2\% compressive (tensile) strain from the unstrained value (Fig. \ref{fig:sinw_thermal_cond}a,b). Similar strain behavior is obtained in SiNWs with W=H up to 5nm, which are not shown here for the sake of brevity. \textit{Thus, ultra-scaled SiNWs show an anisotropic variation in $\kappa_{l}$ under hydrostatic deformation.}

\begin{figure}[bht!]
	\centering
		\includegraphics[width=3.3in,height=1.6in]{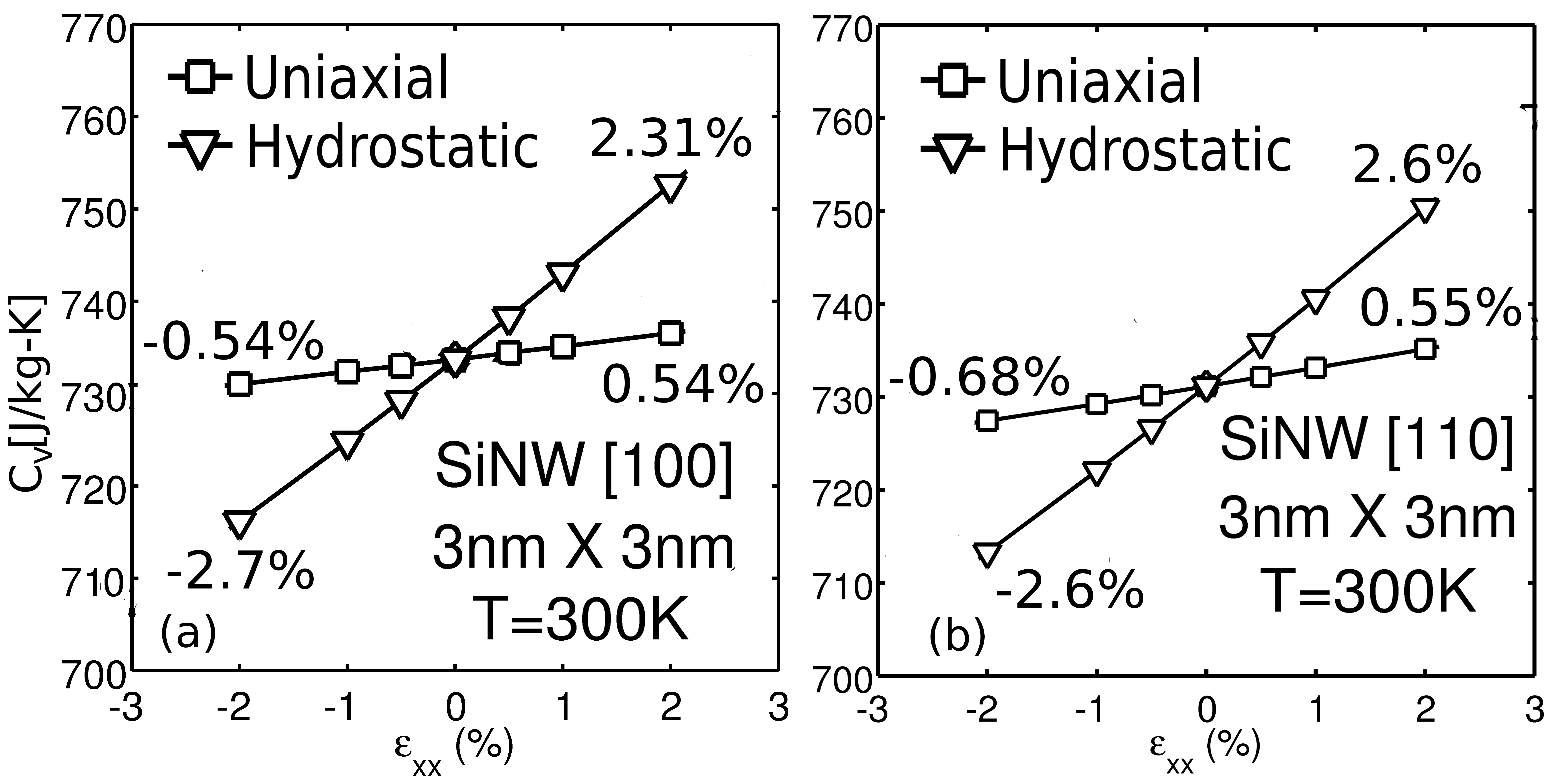}
	\caption{Variation in the $C_{v}$ with stress in 3nm $\times$ 3nm SiNW with (a) [100] and (b) [110] orientation. Two types of stress are applied in these SiNWs, (i) hydrostatic pressure and, (ii) uniaxial stress along the wire axis.}
	\label{fig:sinw_CV_vals}
\end{figure}

\textit{Specific heat ($C_{v}$):} Variation in the phonon dispersion under strain also changes the $C_{v}$ of SiNWs. Compressive (tensile) strain decreases (increases) the $C_{v}$ in both types of SiNWs (Fig. \ref{fig:sinw_CV_vals}a,b)). This variation in $C_{v}$ is similar to the bulk Si behavior as reported in Refs.\cite{strain_effect_2,shape_depend_theory}. Under hydrostatic stress the $C_{v}$ decreases by $\approx 2.7\%$  at 2\% compression and increases to $\approx2.6\%$ at 2\% expansion compared to the unstrained $C_{v}$ value for both the SiNW orientations. Under uniaxial deformation the variation is $\le1\%$ for $\pm$2\% strain for both wire orientations (Fig. \ref{fig:sinw_CV_vals}a,b). \textit{Hence, $C_{v}$ variation is isotropic under strain in ultra-scaled SiNWs.}

\begin{table}[b!]
\centering
\caption{Value of band contribution to $\kappa_{l}$ under strain (Fig. \ref{fig:modes_sinw}). Variation shown from compression (Cm) to tension (Tn) for uniaxial (U) and hydrostatic (H) strain for both SiNW orientations (Or). }
\label{table1}
\begin{tabular}{|c|c|c|c|c|c|}\hline
Strain, [Or] & $\overline{\kappa_{l}(L)}$  & $\overline{\kappa_{l}(M)}$  & $\overline{\kappa_{l}(H)}$& Dom.  & Cm to\\
$E_{i}$ (meV) $\rightarrow$& (0-22) & (22-44) & (44-65) & band & Tn \\
\hline
U, [100] & 36\%-34\% $\downarrow$ & 52\%-50\% $\downarrow$ & 13\%-13\% & L,M &$\downarrow$  \\ 
H, [100] & 32\%-37\% $\uparrow$    & 56\%-45\% $\downarrow$ & 11\%-16\% $\uparrow$ & M & $\downarrow$\\
U, [110] & 49\%-46\% $\downarrow$& 42\%-46\% $\uparrow$ & 10\%-8\% $\downarrow$ & L,H & $\downarrow$  \\
H, [110] & 45\%-48\% $\uparrow$ & 45\%-44\% $\downarrow$ & 7\%-10\% $\uparrow$ & L,H & $\uparrow$ \\
\hline
\end{tabular}   
\end{table}   

\begin{figure}[hbt!]
	\centering
	\includegraphics[width=3.6in,height=2.3in]{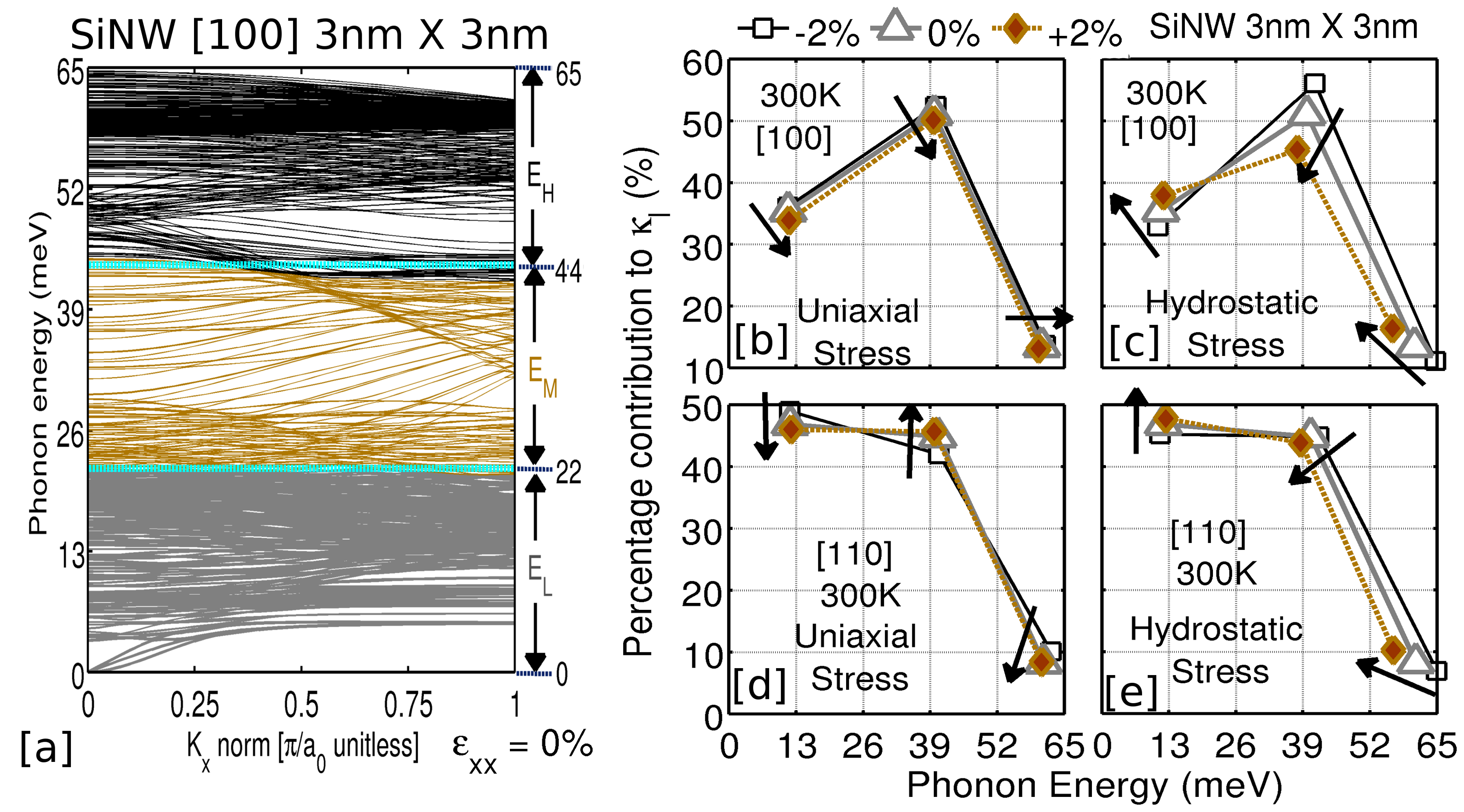}
	\caption{(a) Phonon dispersion of 3nm $\times$ 3nm [100] SiNW with the three energy ranges considered for $\kappa_{l}$ analysis. The percentage contribution to $\kappa_{l}$ from different energy ranges, for [100] SiNW under (b) uniaxial stress, (c) hydrostatic stress and, for [110] SiNW under (d) uniaxial stress and (e) hydrostatic stress. Arrows show the variation in contribution from compression (Cm) to tension (Tn).}
	\label{fig:modes_sinw}
\end{figure}

\textit{Reason for the variation in $\kappa_{l}$:} To understand the variation in $\kappa_{l}$ with strain, the energy dependent contributions are analyzed using Eq. (\ref{eq_kappa}). The entire phonon spectrum is grouped into three energy ranges, (i) $E_{L}$ (`low' bands), (ii) $E_{M}$ (`mid' bands) and, (iii) $E_{H}$ (`high' bands) (see Fig. \ref{fig:modes_sinw} a). These bands show variable contributions under strain which determine the overall effect on $\kappa_{l}$ (Table \ref{table1}).


Under uniaxial strain, in [100] SiNW contribution to $\kappa_{l}$ is mainly from $E_{L}+E_{M}$, which decreases from $\sim$88\% under compression (Cm) to $\sim$84\% under tension (Tn) (Fig. \ref{fig:modes_sinw}b), whereas in [110] SiNW the contribution to $\kappa_{l}$ is mainly from $E_{L}+E_{H}$, which reduces  from $\sim$59\% (Cm) to $\sim$54\% (Tn) (Fig. \ref{fig:modes_sinw}d). Under hydrostatic strain, in [100] SiNW main contribution to $\kappa_{l}$ is from $E_{M}$, which reduces from $\sim$56\% (Cm) to $\sim$45\% (Tn) (Fig. \ref{fig:modes_sinw}c), while in [110] SiNW main contribution to $\kappa_{l}$ is from $E_{L}+E_{H}$, which increases from $\sim$52\% (Cm) to $\sim$58\% (Fig. \ref{fig:modes_sinw}e). These details are also summarized in Table \ref{table1}. \textit{Thus, low/mid energy contribution for [100] SiNW and low/high energy contribution for [110] SiNW explains the anisotropic nature of $\kappa_{l}$. }

\begin{figure}[bht!]
	\centering
		\includegraphics[width=3.0in,height=2.9in]{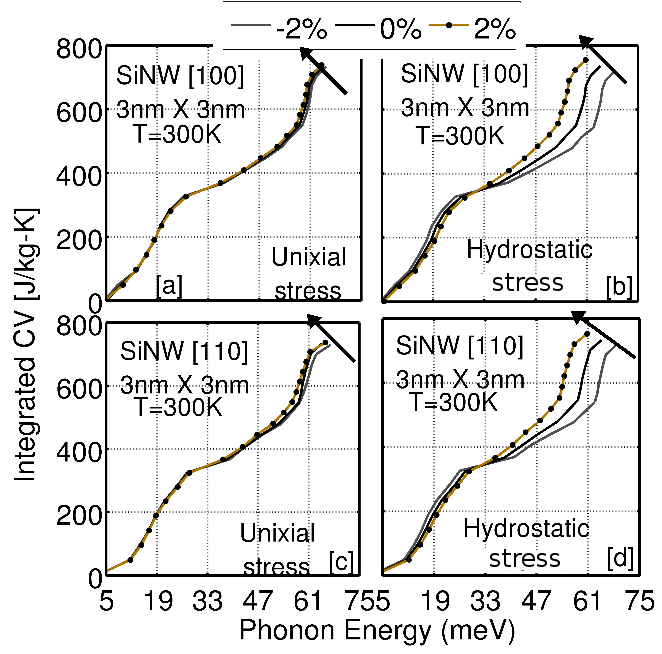}
	\caption{Variation in CV with phonon energy bands under various conditions. 3nm $\times$ 3nm [100] SiNW with (a) uniaxial stress, (b) hydrostatic pressure.
		3nm $\times$ 3nm [110] SiNW with (c) uniaxial stress, (d) hydrostatic pressure. The higher sub-bands show larger variation in CV contribution compared to the lower energy sub-bands. }
	\label{fig:CV_integrated}
\end{figure}

\textit{Reason for $C_{v}$ variation:} Under the action of strain the contribution of the phonon bands in $E_{H}$ range varies considerably as shown in Fig. \ref{fig:CV_integrated}. Under uniaxial stress, both [100] and [110] SiNWs show almost no variation in the $E_{L}$ and $E_{M}$ range, and minute variation in the $E_{H}$ range(Fig. \ref{fig:CV_integrated}a,c). The contribution increases from Cm to Tn in the $E_{H}$ range which governs the $C_{v}$ trend under uniaxial strain (Fig. \ref{fig:sinw_CV_vals}). Under hydrostatic stress, contribution from the $E_{H}$ range increases from Cm to Tn, but in a much larger magnitude compared to uniaxial stress, for both the wire orientations (Fig. \ref{fig:CV_integrated}b,d). This explains the larger variation in $C_{v}$ under hydrostatic stress (Fig. \ref{fig:sinw_CV_vals}). \textit{Thus, the higher energy bands decide the strain behavior of $C_{v}$ under strain, for ultra-scaled SiNWs.}

\textit{Conclusions and Outlook:} The impact of strain on the thermal properties of ultra-scaled SiNWs has been provided. The ballistic thermal conductance behaves an-isotropically under strain, however, $C_{v}$ shows an isotropic behavior. The observed trends in the thermal properties can be understood by the different types of contribution of phonon modes in different energy ranges. The behavior of $C_{v}$ in SiNWs is similar to bulk Si however, $\kappa_{l}$ variation is very different from bulk Si. Under the low stress condition hydrostatic strain can be beneficial in engineering $C_{v}$ and uniaxial stress for engineering $\kappa_{l}$ for cooling lasers. However, high external strain ($|\epsilon|>2\%$ ) field may be needed to engineer $\kappa_{l}$ to improve the efficiency of TE devices \cite{zhang_2005}.

The authors acknowledge financial support from MSD and FCRP under SRC, NRI under MIND, NSF and Purdue University and computational support from nanoHUB.org, an NCN operated and NSF funded project.



%

\newpage

\end{document}